\documentclass[aps,prl,superscriptaddress,twocolumn,showpacs]{revtex4-1}

\usepackage{natbib}
\usepackage{graphicx}
\usepackage[ansinew]{inputenc}
\usepackage{bm}

\newcommand{\unit}[1]{\mathrm {\,#1}}

\newcommand{\etal}{{\em et al.~}}
\newcommand{\kpar}{k_{\|}}
\newcommand{\meff}{m^*}
\newcommand{\Kbar}{$\overline{\text{K}}$}
\newcommand{\Gammabar}{$\overline{\Gamma}$}

\newcommand{\ket}[1]{\ensuremath{|#1\rangle }}
\newcommand{\hpr}[2]{\ensuremath{\langle #1|#2\rangle}}

\begin{document}

\title{Local Gating of an Ir(111) Surface Resonance by Graphene Islands}

\author{S. J. Altenburg}
\email{altenburg@physik.uni-kiel.de}
\affiliation{Institut für Experimentelle und Angewandte Physik, 
Christian-Albrechts-Universität zu Kiel, D-24098 Kiel, Germany}

\author{J. Kröger}
\affiliation{Institut für Physik, Technische Universität Ilmenau, D-98693 
Ilmenau, Germany}

\author{T. O. Wehling}
\author{B. Sachs}
\author{A. I. Lichtenstein}
\affiliation{I.\ Institut für Theoretische Physik, Universität Hamburg, D-20355 Hamburg, Germany}

\author{R. Berndt}
\affiliation{Institut für Experimentelle und Angewandte Physik, 
Christian-Albrechts-Universität zu Kiel, D-24098 Kiel, Germany}

\begin{abstract}
The influence of graphene islands on the electronic structure of the Ir(111) surface is investigated.
Scanning tunneling spectroscopy (STS) indicates the presence of a two-dimensional electron gas with a binding energy of $-160\unit{meV}$ and an effective mass of $-0.18\unit{m_e}$ underneath  single-layer graphene on the Ir(111) surface.
Density functional calculations reveal that the STS features are predominantly due to a holelike surface resonance of the Ir(111) substrate.
Nanometer-sized graphene islands act as local gates, which shift and confine the surface resonance.
\end{abstract}

\pacs{73.20.At, 73.21.Fg, 73.22.Pr}

\maketitle

On a number of metal surfaces, single layers of graphene may be grown.
Owing to their different graphene--metal interactions \cite{Wintterlin20091841}, these substrates may modify the electronic structure of the graphene layer.
For example, the $\pi$ and $\pi^*$ bands of graphene, which give rise to a Dirac cone dispersion in the pristine material, are significantly altered on Ni(111) and Ru(0001) \cite{PhysRevB.50.17487, Himpsel1982L159}, where a gap in the Dirac bands opens.
In contrast, from Ir(111), an almost unchanged band structure of graphene has been reported \cite{art:grapheneMinigapsIr}.  
The modification of the electronic structure of the substrate upon graphene adsorption has hardly been investigated.
From calculations, quenching of the Ni(111) surface state upon graphene adsorption was predicted \cite{dze_11}.
The only experiment related to this issue addressed a spatial variation of the  Ru(0001) $d$ states, which arises from a graphene-induced moiré pattern \cite{mgy_11}. 
This state of affairs is surprising as electronic states at surfaces are sensitive probes of the interaction with adsorbates.

Here, we combine scanning tunneling spectroscopy and density functional theory (DFT) calculations of Ir(111) covered with a single layer of graphene.  
Spectra of the differential conductance ($\text{d}I/\text{d}V$; $I$: current; $V$: sample voltage) of pristine and graphene-covered Ir(111) reveal a holelike surface resonance.
Intriguingly, the most prominent features in the spectra are due to this resonance at the Ir--graphene interface rather than to any states of the graphene layer.
This effect is attributed to the selectivity of the tunneling current for states with small parallel momentum.
Although the electronic structure of graphene is only weakly perturbed by the Ir(111) substrate, graphene shifts the Ir surface resonance by $\approx\,190\unit{meV}$ towards the Fermi level.
As a result, graphene islands act as local gates which confine the surface resonance and induce characteristic standing wave patterns.
The resonance shift and an effective mass $m^*= -0.18\unit{m_e}$ ($\text{m}_{\text{e}}$: free electron mass) determined from these patterns are consistent with DFT results.

\begin{figure}
\includegraphics[width=85mm]{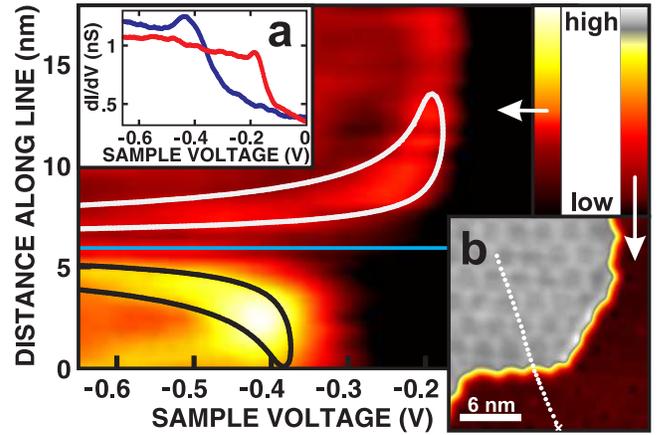}
\caption{Main figure: Spectra of $\text{d}I/\text{d}V/(I/V)$ recorded along a line from Ir onto a graphene island as indicated in Inset (b).
A clear shift of the surface resonance onset occurs.
Contours of constant LDOS calculated using a scattering model (see text) are indicated by black and light gray lines for Ir and graphene, respectively.
A horizontal blue (gray) line indicates the position of the graphene edge.
Inset (a): Spectra of $\text{d}I/\text{d}V$ recorded near the beginning (blue(black)) and end (red(dark gray)) of the line in Inset (b).  The beginning of the line is defined by the cross at the dotted line in Inset (b).
Inset (b): Constant-current STM image of a graphene patch on Ir(111) ($-220\unit{mV}$, $100\unit{pA}$).  
White dots mark positions where $\text{d}I/\text{d}V$ spectra were recorded.  
A white cross denotes zero distance.
The false colors used in the main figure and in Inset (b) are defined in the upper-right-hand corner.}
\label{fig:step}
\end{figure}

Experiments were performed with a home-built scanning tunneling microscope (STM), operated at $5.2\unit{K}$ in an ultrahigh vacuum. 
Ir(111) surfaces were cleaned by cycles of Ar$^+$ bombardment and annealing.
Graphene films were grown by exposing the sample to $\approx 6\times 10^{-4}\unit{Pa}\unit{s}$ of $\text{C}_2\text{H}_4$ at room temperature and subsequent annealing at $\approx 1400\unit{K}$.
This procedure leads to a partial coverage of Ir(111) by highly ordered graphene \cite{art:IrClusters}.
Au tips were prepared by {\it ex-situ} cutting, {\it in-vacuo} heating and Ar$^+$ bombardment. Spectra of $\text{d}I/\text{d}V$ were acquired by a standard lock-in technique (modulation frequency: ~9.1\,{kHz}, modulation amplitude:~10\,{mV$_{\text{rms}}$}) and subsequently normalized by $I/V$ to compensate for the voltage-dependent transmission of the tunneling barrier \cite{art:FeenstraNorm}.
 
Figure \ref{fig:step} shows normalized $\text{d}I/\text{d}V$ data, which were obtained along a slightly curved line [Inset (b) to Fig.\,\ref{fig:step}] crossing a step from bare Ir onto a graphene island. 
Above the bare Ir(111) surface, a steplike decrease in the $\text{d}I/\text{d}V$ signal at $V \approx\,-350\unit{mV}$ [Inset (a) to Fig.\,\ref{fig:step}, blue (black) line] occurs, which is indicative of a holelike surface state or resonance\,\footnote{The large width of the onset is consistent with the fact that the resonance is degenerate with bulk states. It may also be affected by residual adsorbates on Ir(111), which attenuate the signal of the surface resonance in their vicinity. Hence, spectra were obtained as far away from such defects as possible.}.
Such a resonance was previously observed with photoelectron spectroscopy \cite{veen_80,ARPESIr}.
Above the graphene layer, the steplike feature is shifted to $V \approx\,-150\unit{mV}$ [Inset (a) to Fig.\,\ref{fig:step}, red (dark gray) line].
When approaching the edge of the graphene island from either side, the $\text{d}I/\text{d}V$ step moves towards lower voltages and disappears on top of the graphene edge.
This spatial variation of the shift can be explained by scattering at the graphene edge.
The rather strong interaction between graphene edges and the Ir(111) substrate bends the graphene edges towards the metal and leads to the formation of chemical bonds between Ir and C atoms at the island edge \cite{PhysRevLett.103.166101}. 
The Ir(111) surface resonance is considered as a free electron gas with a binding energy $E_0$ and effective mass $\meff$ scattered from a hard-wall potential provided by the graphene edges. 
The spatial variation of the local density of states (LDOS), $\rho_{\text{s}}$, can then be described as
\begin{equation}
\rho_{\text{s}}(E,x)=1-J_0(2\kpar x),
\end{equation} 
where $J_0$ is the zeroth-order Bessel function, $\kpar=\sqrt{2\meff(E-E_0)}/\hbar$ is the parallel momentum, $E$ is the energy, and $x=0$ is the position of the hard-wall potential \cite{avouris:1447,jacklevic,Eigler}. 
Lines in Fig.\,\ref{fig:step} show contours of constant LDOS calculated for $\meff=-0.18\unit{m_e}$ (see confinement analysis below), $E_{0,\,\text{Ir}}=-350\unit{meV}$ (curved black line) and $E_{0,\,\text{gr}}=-160\unit{meV}$ (curved white line), for Ir(111) and graphene-covered Ir(111), respectively.
To match experimental data, a Gaussian broadening of $40\unit{meV}$ was applied.  Further, the energies $E_{0,\,\text{gr}}$ and $E_{0,\,\text{Ir}}$ were chosen to yield a good fit between the calculated first maximum of the oscillation (embraced by contour lines) and the experimental data.
The simple hard-wall model reproduces the curvature and position of the LDOS maxima quite well and yields an energetic shift of the surface resonance between bare and graphene-covered Ir of $\Delta E\approx\,190\unit{meV}$.
The energies fit well to the energies extracted from the single spectra; $E_{0,\,\text{gr}}$ also matches the energy obtained by confinement analysis (see below).
It is important to note that the data do not reveal any particle-hole symmetric counterpart of these confinement features above the Fermi level.
At energies below $\approx\,-650\unit{meV}$ (not shown in Fig.\,\ref{fig:step}) variations in the LDOS with the periodicity of the moiré pattern predominate, probably due to weak periodic potential modulations. 

\begin{figure}
\includegraphics[width=85mm]{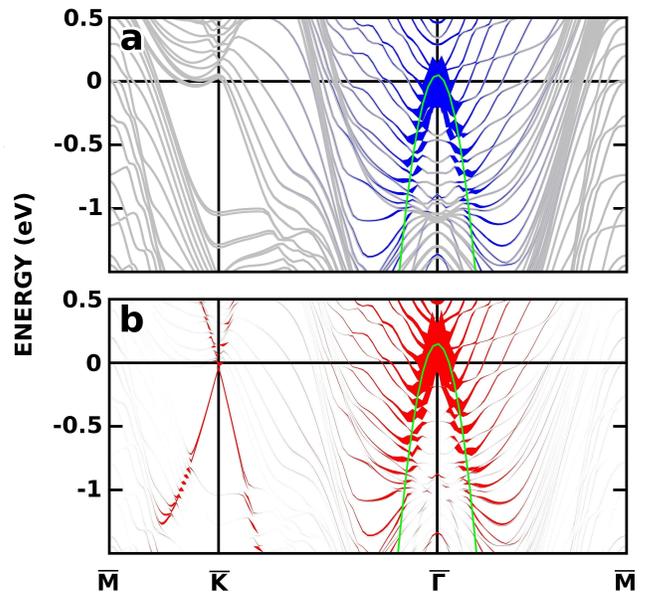}
\caption{
Calculated band structures of (a) pristine and (b) graphene-covered Ir(111).
Light gray lines in (a) show the dispersion of all states of the supercell used.
The contribution of each state to the tunneling current is indicated by the widths of blue (black) and red (dark gray) lines.
On Ir, the current is essentially due to a surface resonance at the center of the surface Brillouin zone (\Gammabar).
On the graphene-covered surface, the resonance is shifted upwards. 
It still carries most of the current, while the Dirac cone around {\Kbar} is less conducting.
Green (gray) parabolas show a fit with an effective mass $\meff\approx-0.17\unit{m_e}$.}
\label{fig:DFT}
\end{figure}

To further support the above model, DFT calculations of the pristine and graphene-covered Ir(111) surface were performed \footnote{The Ir surface was modeled using a slab of 18 layers of Ir atoms and a vacuum gap of $5.5\unit{nm}$.  
A $k$-mesh of $15\times 15\times 1$ points and a kinetic energy cut-off of $400\unit{eV}$ were used.  
To study graphene on Ir(111), the slab was covered with graphene on one side with one of the two C atoms of the graphene unit cell on top of an Ir atom and the graphene lattice constant adjusted to match the Ir lattice. 
Spin-orbit coupling was taken into account to correctly model splitting of surface states \cite{ARPESIr}.}. 
We used the projector-augmented plane wave method \cite{PAWbloechl,PAWkresse}, as implemented in the Vienna \textit{ab initio} simulation package \cite{kresse_vasp}.
Light gray lines in Fig.\,\ref{fig:DFT}(a) show the calculated band structure of an Ir(111) surface.  
Near the Fermi energy of the Ir(111) system, various bands originate from bulk bands with $p$ or $d$ character and there are several surface states around the bulk band gap near {\Kbar} and a surface resonance near \Gammabar.
This resonance is mainly derived from Ir $p_z$ orbitals of the first few atomic layers near the surface.  
In the energy range $-1\unit{eV}<E<0\unit{eV}$ probed in the STM experiments, the calculated dispersion of the surface resonance is approximately parabolic, with an effective mass of $\meff\approx-0.17\unit{m_e}$ [Fig.\,\ref{fig:DFT}(a), green (gray) line].
At the graphene-covered surface [Fig.\,\ref{fig:DFT}(b)], two additional bands derived from the C $p_z$ orbitals occur and form a Dirac cone near {\Kbar} in agreement with previous photoemission \cite{ARPESIr} and DFT \cite{art:grapheneMinigapsIr} studies.
On the graphene-covered surface, the resonance is shifted upwards [Fig.\,\ref{fig:DFT}(b), green (gray) line] by an amount which depends on the distance between the graphene sheet and the topmost Ir layer.
For typical spacings between $0.327$ and $0.362\unit{nm}$ \cite{michelyvdW}, the calculated shift is between $100$ and $200\unit{meV}$, which is consistent with our experimental value of $\approx\,190\unit{meV}$.

To trace back the origin of the resonance shift upon graphene adsorption, calculations were performed in which the graphene C atoms were replaced by chemically fully inert Ne atoms.
As a result, a Ne layer shifts the Ir surface resonance upwards by virtually the same amount as the graphene layer.
As Ne provides no states at the Fermi level which could donate or accept charge from the Ir surface, the upward shift of the resonance is most likely due to a significant Pauli repulsion. 
Nevertheless, Coulomb potential effects, e.\,g., via charge redistribution \cite{michelyvdW}, occur and cannot be disregarded in modeling the full electronic structure of graphene/Ir(111).
The calculations further show a downward shift of the Dirac cone when the graphene sheet is pushed towards the Ir substrate.
This shift cannot be explained by Pauli repulsion and demonstrates that Coulomb potential effects are predominant for the energy of the graphene Dirac point.

To determine the contributions of the various states to the tunneling current, the approach of Tersoff and Hamann \cite{art:Tersoff,*art:Tersoff2} was used and the tip was modeled as an $s$ orbital $\ket{L}$. 
Based on an estimate of the experimental tip--sample distance\,\cite{fnote1} the orbital is placed $0.48\unit{nm}$ above the surface.
The overlap $|\hpr{\Psi_{n,k}}{L}|^2$, where $\ket{\Psi_{n,k}}$ is the wave function of a band $n$ at wave vector $k$, is indicated by the width of the colored bands in Fig.\,\ref{fig:DFT}.
On both surfaces, clean and graphene-covered Ir(111), the main contribution to the current is due to the aforementioned surface resonance.
The current due to the Dirac bands of graphene is significantly smaller.
This may be understood from the parallel momenta $\kpar$ of these states, which affect their decay into vacuum.
The surface resonance is located around {\Gammabar} and thus decays less rapidly than the Dirac cone states near {\Kbar} \footnote{The periodic potential associated with the moiré superlattice of graphene on Ir(111), which is not included in the calculations, may in principle scatter states from {\Kbar} to the center of the Brillouin zone.
However, owing to the large lattice constant of the superstructure, the associated momentum is small and multiple scattering steps are required, which makes this mechanism unlikely.}.
Therefore, the steps in the $\text{d}I/\text{d}V$ spectra may safely be attributed to the (shifted) Ir(111) surface resonance.
This result is also in agreement with the absence of electron--hole symmetry from the experimental spectra.
In recent publications \cite{PhysRevLett.107.236803, doi:10.1021/nn2028105}, scanning tunneling spectroscopy data from graphene on Ir(111) are attributed to tunneling from graphene states.
However, the analyses of Refs.\ \onlinecite{PhysRevLett.107.236803, doi:10.1021/nn2028105}
neglect the substrate electronic states at the Brillouin zone center.
In contrast, the present results show the importance of substrate states at {\Gammabar} which give the dominant contribution to the current in our STM experiments.

\begin{figure}
\includegraphics[width=85mm]{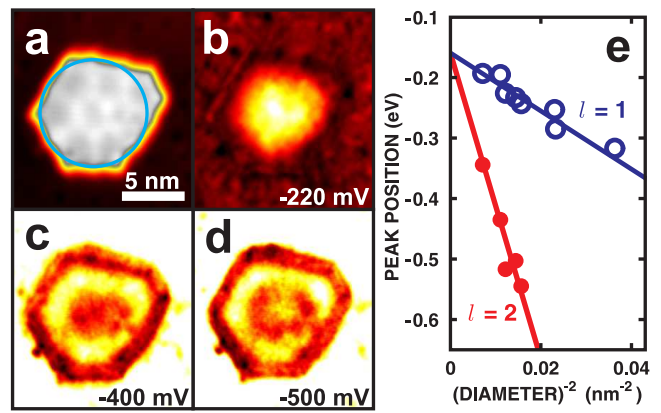}
\caption{(a) Constant-current STM image of a graphene island on Ir(111) ($-220\unit{mV}$, $1\unit{nA}$).  
A blue (light gray) circle indicates the effective island diameter of $8.3\unit{nm}$ used for further analysis.  
(b)--(d) Normalized $\text{d}I/\text{d}V$ maps of the graphene island in (a), recorded at the indicated voltages.
The observed LDOS oscillations evolve as expected for the confinement of an electron gas in a quantum dot.  
(e) Energies of confined states with significant LDOS at the island center ($n=0$) measured from various graphene islands.  
States with $l=1$ (open blue circles) and $l=2$ (filled red circles) are resolved.  
Lines show a fit according to Eq.\,(\ref{eqn:Enl}) with $E_0=-160\unit{meV}$ and $\meff=-0.18\unit{m_e}$, which are in good agreement with values calculated within DFT\@.} 	
\label{fig:maps}
\end{figure}

In addition to scattering at their edges, graphene islands lead to confinement of the hole states.  
Figure\,\ref{fig:maps} shows an STM image of a graphene island along with normalized $\text{d}I/\text{d}V$ maps recorded at constant current.
At increasingly negative sample bias [Figs.\,\ref{fig:maps}(b)--(d)] the
pattern inside the island evolves from a central maximum over a ring to a ring with a central maximum, as expected for confined states with zero, one, and two nodes, respectively.
For a more detailed analysis, we model the island by a circular quantum dot with hard walls. 
The eigenenergies $E_{n,l}$ of a confined electron gas are \cite{platt:1448}
\begin{equation}
	E_{n,l} = E_0 + \frac{2\hbar^2u_{n,l}^2}{\meff d^2},
	\label{eqn:Enl}
\end{equation}
where $u_{n,l}$ is the $l$th root of the $n$th-order Bessel function $J_n$ and $d$ is the island diameter.
Figure \ref{fig:maps}(e) displays the energies of the first two resonances which exhibit an LDOS maximum at the island center ($E_{0,1}$ and $E_{0,2}$) evaluated from spatially resolved $\text{d}I/\text{d}V$ spectra of 8 nearly circular islands with diameters between $5$ and $12\unit{nm}$.
The effective island diameters $d$ were determined from an inscribed circle, which touches the island boundaries at the midpoint of the step edge [Fig.\,\ref{fig:maps}(a), blue (light gray) line].
Energies calculated according to Eq.\,(\ref{eqn:Enl}) with  $E_0=-160\unit{meV}$ and $\meff=-0.18\unit{m_e}$ [Fig.\,\ref{fig:maps}(e), lines] match the experimental data very well.

\begin{figure}
\includegraphics[width=85mm]{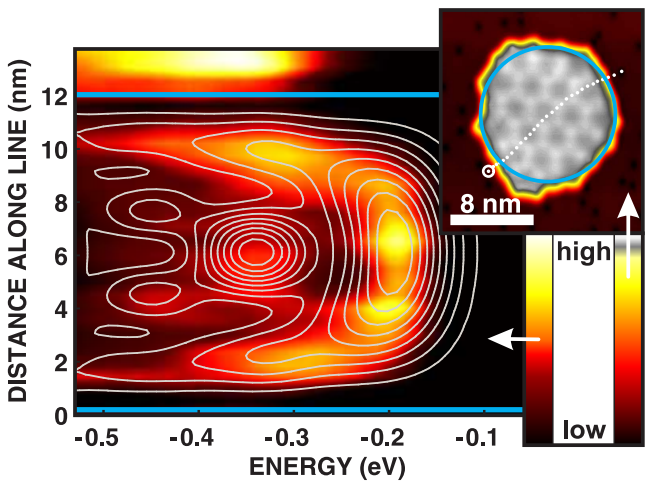}
\caption{Spatially resolved normalized $\text{d}I/\text{d}V$ spectra recorded along a line through a  graphene island (inset).
The experimental distance scale was recalibrated to account for the small curvature of the measurement path.
Light gray lines mark contours of constant LDOS calculated for the circular island which is indicated in the inset.
Blue (gray) lines indicate the boundaries of the model island.
Inset: Constant-current STM image of the graphene island ($-220\unit{mV}$, $100\unit{pA}$).
White dots indicate positions where spectra of $\text{d}I/\text{d}V$ were recorded.
A white ring denotes zero distance.
An inscribed blue (gray) circle shows the effective island diameter of $11.8\unit{nm}$ used in modeling.}
\label{fig:largeIsland}
\end{figure}

As a final test of the model spatially resolved $\text{d}I/\text{d}V$ spectra from a roundish island are compared with the calculated LDOS\@. 
The LDOS, $\rho(E,x)$, of the surface resonance confined to a disk is
\begin{equation}
	\rho(E,x) = \sum_{n,l}
	|\Psi_{n,l}|^2
	\exp\left[
	-\frac{1}{2}\left(
	\frac{E_{n,l}-E}{\delta E}
	\right)^2
	\right].  
	\label{eqn:rhoIsland}
\end{equation}
$\Psi_{n,l}$ are solutions to the Schrödinger equation as described by Platt \etal \cite{platt:1448},
\begin{equation}
\Psi_{n,l}(r,\varphi) = J_n\left(
u_{n,l}\frac{2r}{d}
\right)
\exp(\text{i}n\varphi).
\label{eqn:PsiIsland}
\end{equation}
$\delta E$ is a Gaussian broadening reflecting a finite lifetime of the states.
While the broadening may depend on energy \cite{PhysRevB.71.155417}, the constant broadening $\delta E=40\unit{meV}$ assumed here is sufficient to match the experimental observations.
Figure \ref{fig:largeIsland} shows a series of 29 normalized $\text{d}I/\text{d}V$ spectra measured along a line across a graphene island (inset).
Light gray contour lines show the calculated LDOS using the measured diameter of $11.8\unit{nm}$ (blue (gray) horizontal lines).
The qualitative agreement of the theoretical and experimental data is further evidence that graphene islands confine the Ir(111) resonance.

In conclusion, van der Waals-bonded graphene on Ir(111) induces a pronounced shift in the Ir(111) surface resonance.  The disappearance of the resonance at graphene edges indicates the covalent carbon-metal interaction, which acts as a hard-wall potential for scattering of resonance electrons.  Nanometer-sized graphene flakes can therefore confine quasi-two-dimensional electron gases to artificial quantum dots.

Funding by the Deut\-sche For\-schungs\-ge\-mein\-schaft via SPP 1459 and SFB 668, and the Schles\-wig-Hol\-stein-Fonds, as well as computer time at HLRN, are acknowledged.

Note added in proof: Results concerning confined electronic states in graphene islands on Ir(111) \cite{PhysRevLett.108.046801} have been published after submission of this manuscript. Ref.\ \onlinecite{PhysRevLett.108.046801} attributes the confined states to either graphene states or the scattered Ir(111) surface resonance, depending on the graphene island size. This is in contrast to our interpretation that the surface resonance predominates the STM data for any island size.

During the refereeing process an experimental observation of a graphene-induced shift of the Ir(111) surface resonance has been reported~\cite{PhysRevLett.108.066804}.

\bibliography{cites2}

\begin{thebibliography}{32}
\bibitem [{fno()}]{fnote1}%
  \BibitemOpen
  \href@noop {} {}\bibinfo {note} {To estimate the tip-sample distance $z$ the
  tip was approached to the surface and the current $I$ was simultaneously
  measured. A drastic deviation of $I(z)$ from an exponential signals the
  formation of a contact \cite{art:atomTransfer,sal_10}. Assuming a typical
  bond length of $z \approx 0.2\unit{nm}$ at contact, we find $z\approx
  0.5\unit{nm}$ at typical tunneling conditions.}\BibitemShut {Stop}%
\end{thebibliography}
\end{document}